\documentstyle[12pt]{article}
\newcommand{\beq}{\begin{equation}}
\newcommand{\eeq}{\end{equation}}
\newcommand{\ber}{\begin{eqnarray}}
\newcommand{\eer}{\end{eqnarray}}
\begin{document}
\title{Factorization and Effective Theories}

\author{
{\it Ugo Aglietti, Guido Corb\`o}\\
\\
Dipartimento di Fisica, Universit\` a di Roma \lq La Sapienza\rq \\
INFN, Sezione di Roma I, P.le A. Moro 2, 00185 Roma, Italy.}        
\date{}
\maketitle
\begin{abstract}

We prove factorization in the decay $B~\rightarrow~D^{(*)}+jet$
using the Large Energy Effective Theory.
The proof is non perturbative,
does not require any gauge fixing and is exact in the limit of a very narrow
jet.
On the other hand, it is shown that the Large Energy Effective Theory 
is unable to consistently
describe completely exclusive processes such as 
for example $B~\rightarrow~D^{(*)}+\pi$ due to an oversimplification of
transverse momentum dynamics. Therefore
we present
a variant of the Large Energy Effective Theory, i.e. a new
effective theory for massless particles
which properly takes into account transverse degrees of freedom
and is the natural framework to study exclusive
non-leptonic decays.

\end{abstract}
\newpage

We consider the decay
\beq\label{uno}
B~\rightarrow~D^{(*)}+jet
\eeq
in the limit $m_b-m_c \rightarrow \infty$.
The process is described by the following 
four-point correlation function:
\beq\label{due}
\langle 0\mid T~L(z)~D(y)~{\cal{H}}_{W}(0)~B^{\dagger}(x)
\mid 0\rangle
\eeq
taking the asymptotic limits
\beq\label{tre}
t_x~\rightarrow~-\infty,~~~~~
t_y~\rightarrow~+\infty,~~~~~
t_z~\rightarrow~+\infty.
\eeq
$B(x)=\overline{q}(x)\Gamma_B b(x)$, $D(x)=\overline{q}(x)\Gamma_D c(x)$ and
$L(x)=\overline{u}(x)\Gamma_L d(x)$ are
interpolating fields for the $B$, $D^{(*)}$ and the
jet containing {\it up} and {\it down} quarks respectively;
$\Gamma_{B,D,L}$ are suitable matrices in Dirac space and
${\cal{H}}_W(x)$ is the effective non-leptonic weak Hamiltonian which
may be written as
\beq\label{quattro}
{\cal{H}}_W(x)~=~\frac{G_F}{\sqrt{2}}~\Big[C_1 O_1(x)~+~C_8
O_8(x)\Big].
\eeq
$G_F$ is the Fermi constant and $O_1(x)$, $O_8(x)$ are
local four-fermion operators which
can be written as

\beq\label{cinque}
O_i(x) = g^{\mu \nu}
J_{i\mu}(x)~j_{i\nu}(x)
\eeq
where
\ber
J_{i\mu}(x) &=& \overline{c}(x)~\gamma_{\mu L}\xi_i~b(x)
\nonumber\\
j_{i\mu}(x) &=& \overline{d}(x)~\gamma_{\mu L}\xi_i~u(x)
\eer
are
singlet and octet currents in colour space.
$\gamma_{L}^{\mu}=\gamma^{\mu}(1-\gamma_5)$,
$\xi_i=1,t^a$ for $i=1$ and 8 respectively.
$C_1$, $C_8$ are Wilson coefficients resumming hard
gluon effects of the form $\alpha_S^n\log^k(m_W^2/m_b^2),$
$k\leq n$ \cite{refuno}.

Performing the functional integration over the fermionic fields we have:

$$
\langle 0\mid T~L(z)~D(y)~{\cal{H}}_{W}(0)~B^{\dagger}(x)
\mid 0\rangle
$$
\beq\label{cinqueuno}
=\frac{1}{N}\int {\cal{D}}A_{\mu}~e^{iS_{eff}[A_{\mu}]}\langle 0\mid
T~L(z)~D(y)~{\cal{H}}_{W}(0)~B^{\dagger}(x)
\mid 0\rangle_A
\eeq
where
$$
S_{eff}[A_{\mu}] = S_{YM}[A_{\mu}] + n~ln~ det~ [~i\hat{D}~(A_{\mu})~]
$$
is an effective action for the gauge field involving fermion loops.
$n$ is the number of flavours and
$$
N = \int {\cal{D}}A_{\mu}~e^{iS_{eff}[A_{\mu}]}.
$$

According to the Wick theorem we write for the correlation in a given gauge
field $A_{\mu}$:
$$
\langle 0\mid T~L(z)~D(y)~{\cal{H}}_{W}(0)~B^{\dagger}(x)
\mid 0\rangle_A
$$
$$
=\frac{G_F}{\sqrt{2}}\sum_{i=1,8}~C_i~
\langle 0\mid T~D(y)J_{i\mu}(0)
B^{\dagger}(x) \mid 0 \rangle_A
~\cdot \langle 0 \mid T~L(z)j_{i}~^{\nu}(0) \mid 0 \rangle_A
$$
$$                
=\frac{G_F}{\sqrt{2}}\sum_{i=1,8}~C_i
Tr\Big[~iS_c(y\mid 0;A_{\mu})~\gamma_{\mu L}\xi_i~iS_b(0\mid x;A_{\mu})~
\tilde{\Gamma}_B~iS_{s}(x\mid y;A_{\mu})~\Gamma_D~\Big]
$$
\beq\label{undici}
\cdot
Tr\Big[~iS_d(z\mid 0;A_{\mu})~\gamma^{\mu}_L\xi_i~
iS_u(0\mid z;A_{\mu})~\Gamma_L~\Big]
\eeq
where $\tilde{\Gamma}_B= \gamma_0 \Gamma^{\dagger}_B \gamma_0$.
We have taken
a valence strange quark $(q=s)$ to avoid unimportant contractions.
Each contribution involves the product
of two separate `heavy' and `light'
fermionic traces.

Now we replace the {\it down} and {\it up} propagators with
the particle and antiparticle propagators of the 
Large Energy Effective Theory \cite{refdue} ($LEET$) respectively: 
\ber\label{dieci}
iS_d(z\mid 0;A_{\mu})&\rightarrow &iS_0(z)~P\exp
\Bigg[~ig\int_0^{z}dx_{\mu}A^{\mu}(x)~\Bigg]
\nonumber\\
iS_u(0\mid z;A_{\mu})&\rightarrow &iS_0(z)~P\exp
\Bigg[~ig\int_z^{0}dx_{\mu}A^{\mu}(x)~\Bigg]
\eer
where $iS_0(x)$ is the free $LEET$ propagator,
\beq\label{nove}
iS_0(x)~=~\frac{ \hat{n} }{2}~\int_0^{\infty}
d\tau~\delta^{(4)}(x-n\tau)~
=~\frac{\hat{n}}{2}~\theta(t)~\frac{\delta^{(3)}(\vec{x}-\vec{u}t)}{n_0}
\eeq

For the `light' trace we have:
$$         
Tr\Big[~iS_d(z\mid 0;A_{\mu})~\gamma^{\mu}_L~iS_u(0\mid z;A_{\mu})~
\Gamma_L~\Big]
$$
\beq\label{tredici}
=~\frac{\theta(t)}{n_0^2}~
\delta^{(3)}_{\Lambda}(\vec{x}=0)~
Tr_{spin}\Bigg[\frac{\hat{n}}{2}~\Gamma_L~
\frac{\hat{n}}{2}~\gamma^{\mu}_L\Bigg]~Tr_{col}.
\eeq
$\delta^{(3)}_{\Lambda}(\vec{x})$ is a regularized $\delta$-function
\beq\label{undicib}
\delta^{(3)}_{\Lambda}(\vec{x})~=~\int^{\Lambda} \frac{d^3 k}{(2\pi)^3}~
e^{i\vec{k}\cdot\vec{x}}~
=~\frac{\Lambda^3}{6\pi^2}~\delta_{\vec{x},0}
\eeq
and
\beq\label{dodici}
Tr_{col} = Tr\Bigg[ P~e^{i\int_0^z A_{\mu}~dx^{\mu}}~\xi_i~
         P~e^{i\int_z^o A_{\mu}~dx^{\mu}}~\Bigg]~
=~3\delta_{i,1}
\eeq
where $\delta_{i,j}$ is the Kronecker symbol.

We see that
the term with $i=8$ does not contribute to the
correlation. 
The physical reason of this result is that a quark and an antiquark
in the same point in a colour
singlet state cannot emit a gluon and go into an octet state:
the dipole field strength is zero.
The only contribution comes from the term with $i=1$ (i.e. from $O_1$
only). Here we reach the main point: 
the P-lines are one the inverse of the other
so that the dynamics of the light pair is in fact {\it independent} 
of the gauge
field and consequently on the heavy system dynamics.
The functional integration over $A_{\mu}$ involves 
the heavy trace only and we have factorization of the correlator 
\cite{reftre}:
$$
\langle 0\mid T~L(z)~D(y)~{\cal{H}}_{W}(0)~B^{\dagger}(x)
\mid 0\rangle
$$
\beq\label{dodicib}
=\frac{G_F}{\sqrt{2}}~C_1~\langle 0 \mid T~D(y)J^{\mu}_1(0)B^{\dagger}(x) \mid
0\rangle\langle 0\mid T~L(z)j_{1\mu}(0) \mid 0 \rangle
\eeq

The proof we have given
is non perturbative, does not require any gauge fixing 
and is exact in the limit of a very narrow jet 
(an extended version of this analysis is given in ref. \cite{refquattro}).

There are however corrections to factorization in the seminclusive
process (\ref{uno}) related to the fact that the jet formed by the
light $ud$ pair is not infinitely narrow.
Let us assume a jet angular width
\beq\label{quindici}
2~\delta~\ll~1.
\eeq
This implies that the light pair can be emitted
with a relative angle up to
$2~\delta$, so that the light-like vectors $n$ and $n'$ of the
quark and the antiquark can be written,
up to first order, as:
\beq\label{sedici}
n~=~(1;\delta,0,1),~~~~~~~~~~n'~=~(1;-\delta,0,1).
\eeq

We have taken the relative motion
of the pair in the $x$ direction and $n^2=n'^2=\delta^2\sim0$.
The two light quarks are created in the origin and
reach two different points which we call
$z$ and $z'$. The correlation describing the semi-inclusive decay can now be
written
\beq\label{diciassette}
\langle 0\mid T~F(z,z')~D(y)~{\cal{H}}_W(0)~
B^{\dagger}(x) \mid 0 \rangle
\eeq
$F(z,z')$ is a bilinear operator which, by covariance, may be written
as

\beq\label{diciotto}
F(z,z')~=~\overline{Q}_{n'}(z')~\Gamma_L~
P\exp\Bigg[ig\int_{z}^{z'} dx_{\mu}A^{\mu}(x) \Bigg]~Q_{n}(z)
\eeq

There is an ambiguity (of non-perturbative kind) in the choice of the path
connecting the point $z$ with $z'$. We consider small angles of emission of
the quarks and we simply take the small segment joining $z$
with $z'$.

The weak hamiltonian now contains two $LEET$ fields
with different velocities $n$ and $n'$, so that the operators
$O_1$ and $O_8$ in eq.(\ref{quattro})  are of the form
\beq\label{diciannove}
O_i(x)~=~g^{\mu \nu}J_{i\mu}(x)\overline{Q}_{n}(x)~\gamma_{\nu L}\xi_i~Q_{n'}
(x)
\eeq

We have:
$$
\langle 0\mid T~F(z,z')~D(y)~{\cal{H}}_W(0)~
B^{\dagger}(x) \mid 0 \rangle_A
$$
$$
=\frac{G_F}{\sqrt{2}}\sum_{i=1,8}~C_i~
\langle 0\mid T~D(y) J_{i\mu}(0) B^{\dagger}(x) \mid 0 \rangle_A 
$$
$$
\cdot~
\langle 0\mid T~F(z,z') \overline{Q}_{n}(0)~\gamma_L^{~\mu}\xi_i~Q_{n'}
(0)\mid 0 \rangle_A 
$$
$$
=\frac{G_F}{\sqrt{2}}\sum_{i=1,8}~C_i~ 
Tr \Big[~iS_c(y\mid 0;A_{\mu})~\gamma_{\mu L}\xi_i~iS_b(0\mid x;A_{\mu})~
\tilde{\Gamma}_B~
iS_{s}(x\mid y;A_{\mu})~\Gamma_D~\Big]
$$
$$
\cdot~
Tr \Big[iS_n(z\mid 0;A_{\mu})~\gamma^{\mu}_L\xi_i~
iS_{n'}(0\mid z';A_{\mu})~\Gamma_L~P(z'\mid z;A_{\mu})~\Big]
$$
$$
=\frac{G_F}{\sqrt{2}}\sum_{i=1,8}~C_i~
Tr\Big[~iS_c(y\mid 0;A_{\mu})~\gamma_{\mu L}\xi_i~iS_b(0\mid x;A_{\mu})~
\tilde{\Gamma}_B~
iS_{s}(x\mid y;A_{\mu})~\Gamma_D~\Big]
$$
$$
\cdot~
Tr \Big[~P(z\mid 0)~\xi_i~P(0\mid z')~P(z'\mid z)~\Big]
$$
\beq\label{venti}
\cdot~\theta(t_z)~\theta(t_z')~
\frac{ \delta^{(3)}(\vec{z}-\vec{u}t_z) }{n_0}~
\frac{ \delta^{(3)}(\vec{z'}-\vec{u}'t_z') }{n_0'}~ Tr\Bigg[~
\frac{\hat{n}}{2}\gamma^{\mu}_L~\frac{\hat{n'}}{2}\Gamma_L~\Bigg]~
\eeq
In the last member we have replaced $LEET$ propagators for the light quarks 
and a more compact notation $P(y\mid x)$ for a 
P-line joining $x$ with $y$ has been introduced.

Unlike the previous case in which $n'=n$, the P-lines do not cancel
each other any more. There is a 
non trivial dependence of the light trace
on the gauge field and light quark dynamics
is represented by
a Wilson loop along
a thin
triangular path with vertices in the origin and in the points $z=n\tau$,
$z'=n'\tau'$ with some selected value for $\tau$ and $\tau'$.
\bigskip

As noted in ref.\cite{refcinque}, the $LEET$ is affected
by an instability problem.
Let us briefly review this phenomenon, considering
the collision between an effective quark $Q$ with (residual)
momentum $k$ and a massless quark $q$ of momentum $p$ described by the 
Dirac theory.
Energy-momentum conservation gives:
\ber\label{ventuno}
\mid\vec{p}\mid + \vec{u}\cdot\vec{k}&=& \mid\vec{p'}\mid
+\vec{u}\cdot\vec{k'}
\nonumber\\
\vec{p}+\vec{k}&=&\vec{p'}+\vec{k}'.
\eer

Let us assume that $p+k$ is a time-like vector, $(p+k)^2>0$.
In the COM frame, $\vec{p}+\vec{k}=0$,
with $\vec{n}$ oriented along the $+z$ axis, we have
\beq\label{ventidue}
     p  (1-\cos\theta)~=~p' (1-\cos\theta')~=~{\cal{E}}
\eeq
where $\theta$ is the angle between the quark 3-momentum and $\vec{n}$.
The prime denotes final state quantities and
\beq\label{ventitre}
{\cal{E}}~=~\sqrt{(p+k)^2}~\ll~E
\eeq
is the total energy of the system in the $LEET$;
it is expected to be of the order $\Lambda_{QCD}$ in $QCD$ applications,
while $E$ is the hard scale of the process.
The instability originates because
\beq\label{ventiquattro}
p'~=~\frac{{\cal{E}}}{1-\cos\theta'}~\rightarrow~\infty
\eeq
when
\beq\label{venticinque}
\theta'~\rightarrow~0.
\eeq
It is related to the emission of particles in the forward direction
$\vec{n}$, the flight direction of $Q$.

Let us assume now a finite angular resolution $\delta>0$ of the
detectors \cite{refsei}. This implies that $Q$ cannot be distinguished from
almost-collinear partons. We consider a cone of half-opening angle
$\delta$ with the axis along $\vec{n}$.
We have that $q$ is observed as an individual particle
if it is emitted in the final state outside the cone,
\beq\label{ventisei}
\theta'~>~\delta.
\eeq
In this case the energy is bounded by
\beq\label{ventisette}
p'~<~ \frac{{\cal{E}}}{1-\cos\delta}~<~\infty,
\eeq
and cannot diverge anymore. Therefore,
the divergence (\ref{ventiquattro}) does not occur for an observable parton.
On the other hand, if the final parton is inside the cone,
\beq\label{ventotto}
\theta'~<~\delta,
\eeq
the finite angular resolution makes impossible to
detect $q$ and $Q$ as separated particles and to measure
their individual energies. A single particle (jet) is observed
with the sum of the parton energies
\beq\label{ventinove}
p'+\epsilon'~=~\frac{{\cal{E}}}{1-\cos\theta'}
-\cos\theta'\frac{{\cal{E}}}{1-\cos\theta'}~=~{\cal{E}}~<~\infty
\eeq
The individual energies are separately divergent but the sum
is finite (small) by assumption (it equals the initial energy).
The instability is therefore eliminated by the angular separation
requirement.
\bigskip

We now show that the $LEET$ is not adequate to describe
{\it exclusive} channels such as, for example, the decay
\beq\label{quattordici}
B \rightarrow D^{(*)} + \pi.
\eeq

Because of the vanishing of gauge interactions (see eq. (\ref{dodici})),
the light quarks
do not have interactions among themselves. This implies that
the light quarks cannot build up a bound state.
A given {\it exclusive} channel as (\ref{quattordici}) cannot be
selected {\it in principle} from the correlation.

In general the infrared properties of $QCD$ are not reproduced by the
$LEET$, as shown by the following perturbative calculation.

Let us look at the propagator of a light meson (such as a $\pi$)
\beq\label{trenta}
C(x) = \langle 0 \mid T \pi(x)~\pi(0) \mid 0 \rangle
\eeq
where $\pi(x)~=~\overline{u}(x)\Gamma d(x)$.

Its Fourier transform, computed in lowest order $QCD$
perturbation theory, is
\footnote{We neglect the numerator structure of the amplitude
because we are looking only at infrared singularities,
which appear as zeros of the denominator.}:
\ber\label{trentuno}
 C_F &\sim&\int d^4
k~\frac{1}{(xP+k)^2+i\epsilon}~\frac{1}{((1-x)P-k)^2+i\epsilon}
\nonumber\\
&\sim &
\int d^4k
\frac{1}{k_0-k_z+k^2/(2xE)+i\epsilon}
\frac{1}{k_0-k_z-k^2/(2(1-x)E)-i\epsilon}
\nonumber\\
&\sim &\int \frac{d^2 k_T ~ d k_+ ~ d k_-}
{(k_-+k_+k_-/(2xE)-k_T^2/(2xE)+i\epsilon)}~\cdot
\nonumber\\
&&~~~\cdot~\frac{1}{(k_--k_+k_-/(2(1-x)E)+k_T^2/(2(1-x)E)-i\epsilon)}
\eer
where we have taken an external momentum $P=En$
with $n=(1;0,0,1)$ and $E>0$.
The variable $x$ represents the quark momentum fraction in the
infinite momentum frame ($0\leq x \leq 1$)
\footnote{We assume that the quark momentum
distribution in the meson $q(x)$ is not singularly peaked at the endpoints
$x=0,1$, so that $xE$ and $(1-x)E$ can always be considered
as large energies. A physical justification of this assumption
comes from an expected Sudakov suppression of the elastic region.}.
$k_+=k_0+k_z$ and $k_-=k_0-k_z$ are the usual light-cone variables
and $k_T^2=\vec{k}_T^2$.
The poles in the $k_-$-plane are located at
\beq\label{trentadue}
k_-~=~\frac{k_T^2/(2xE)-i\epsilon}{1+k_+/(2xE)},~~~~~
k_-~=~\frac{-k_T^2/(2(1-x)E)+i\epsilon}{1-k_+/(2(1-x)E)}
\eeq

Assuming a cutoff $\Lambda$ on $k_+$ satisfying $\Lambda\ll E$, the
integral is approximated by
\beq\label{trentatre}
C_F\sim\int d^2 k_T\int d k_+ \int d k_-
\frac{1}{k_- - k_T^2/(2xE)+i\epsilon}~
\frac{1}{k_-+k_T^2/(2(1-x)E)-i\epsilon}
\eeq

There is a pinching of the poles in the $k_-$-plane for $k_T=0$:
in other words, the integration contour is trapped
between two poles which coalesce in the limit $k_T\rightarrow 0$.
The integral is logarithmically divergent
\footnote{The infrared logarithmic singularity originates because the integral
(\ref{trentuno}) does not contain any scale ($P^2=0$), so it is of the
form $\int
d^4k/(k^2)^2$, as can be seen explicitly introducing a Feynman
parameter.} with $\epsilon$:
\beq\label{trentaquattro}
C_F~\sim~\int \frac{dk_T^2}{k_T^2-i\epsilon}~\sim~\log\frac{1}{\epsilon}.
\eeq

The effective theory amplitude is obtained taking the limit
$E\rightarrow\infty$ in the integrand:
\ber\label{trentacinque}
C_E(r=0)&\sim&\int d^2 k_T \int d k_0 d k_z~
\frac{1}{k_0-k_z+i\epsilon}~\frac{1}{k_0-k_z-i\epsilon}
\nonumber\\
&\sim&\int d^2 k_T~\int d k_+ \int d k_-
\frac{1}{k_-+i\epsilon}~\frac{1}{k_--i\epsilon}
\eer
where $r$ is the meson residual momentum, which we set to zero.

The integrations over the transverse momentum and over $k_+$ give
a cubic ultraviolet divergence (compare with eq. (\ref{undicib})).
As for the infrared, we see that
the integral over $k_-$ involves a pinch singularity
due to the infinitesimally close poles at $k_-=\pm i\epsilon$.
We note that pinching occurs in the whole transverse momentum space,
while it occurs only for $k_T\rightarrow 0$ in the full
theory.
In other words, infrared divergences in $QCD$ are regulated by
${\vec{k}_T}^2/2E$ while they are regulated by $i\epsilon$ in the
$LEET$.
Integrating over $k_-$ we pick up a $1/\epsilon$, i.e. a
linearly divergent contribution:
\beq\label{trentasei}
C_E\sim \frac{1}{\epsilon}
\eeq

Thus the infrared behaviour of the full theory is not reproduced by the
$LEET$:
a logarithmic infrared singularity in $QCD$ is in fact converted
into a linear (i.e. much stronger) singularity in the effective theory.
This is related to the intrinsic `one-dimensionality' of
the $LEET$ which misses altogether
transverse momentum dynamics
(responsible of bound states
existence).

We can ask ourselves whether it is possible to modify the $LEET$ in such
a way to account for transverse degrees of freedom.
We propose to include
the leading kinetic correction into the propagator:
\beq\label{trentasette}
iS(k)~=~\frac{\hat{n}}{2}~
\frac{i}{n\cdot k-{\vec{k}_T}^2/2E~+i\epsilon}
\eeq
where we have taken $n^{\mu}=(1;0,0,1),~k_T^{\mu}=(0;\vec{k}_T,0)$
\footnote{
We can give a Lorentz invariant representation of the
transverse momentum $k_T$ with the Sudakov basis.
If we define a second light-like vector $\eta$ such that
$n\cdot \eta=2$ (in the usual frame, $\eta=(1;0,0,-1)$),
we have $n\cdot k_T=\eta\cdot k_T=0$ and
$k_T^2=-\vec{k}_T^2=k^2-n\cdot k~\eta\cdot k$.}.
Writing eq.(\ref{trentasette}) we follow an idea of `minimal
correction' of the $LEET$ pathologies; we neglect for example
the term $\hat{k}$ in the numerator of the full propagator,
so that the spin structure is factorized.

We may call this new effective theory `Modified Large
Energy Effective Theory', $\overline{LEET}$ for short.
Note that, unlike the $LEET$ case, the hard scale $E$
is still present in the theory, i.e. it cannot be completely
removed.

The problem of pinch singularities previously discussed
is solved replacing $LEET$ propagators with $\overline{LEET}$
propagators: the meson propagator in the $\overline{LEET}$
has the form (\ref{trentatre}) so that pinch singularities
occur only for $\vec{k}_T=0$ instead of in the whole transverse
momentum space.
The $\overline{LEET}$ amplitude
coincides with the full theory
amplitude (\ref{trentuno}) for $k_+\ll E$ and has the same infrared
behaviour.

The propagator is given as a function of time and spatial momentum
by
\beq\label{trentotto}
iS(t,\vec{k})~=~\frac{\hat{n}}{2}~\theta(t)~
\exp\Bigg[-ik_Zt-i\frac{k_T^2~t}{2E}~\Bigg]
\eeq
and in configuration space by
\ber\label{trentanove}
iS(t,\vec{x})&=&\int
\frac{d^3k}{(2\pi)^3}~e^{i\vec{k}\cdot\vec{x}}~
               iS(t,\vec{k})
\nonumber\\
&=&\frac{\hat{n}}{2}~
\theta(t)~\delta(z-t)~\frac{E}{2\pi it}~e^{iE b^2/(2t)}
\nonumber\\
&=&iS^{LEET}(x)~\frac{E}{2\pi i t}e^{i E b^2/(2t)}
\eer
where $b=\mid\vec{x}_T\mid$ is the impact parameter.
The effect of the transverse momentum term is factorized
and produces a diffusion in the impact parameter space:
the factor in the last line of eq.(\ref{trentanove}) represents
a gaussian process after analytic continuation
$t_M=-it_E$.
At large times, we have:
\beq\label{quaranta}
S(t,\vec{x})~\simeq~\frac{\hat{n}}{2}~
\theta(t)~\delta(z-t)~\frac{E}{2\pi it}.
\eeq
We see that there is a diffusion normal to the classical particle
trajectory $z=t$ produced by transverse momentum fluctuations,
which is instead absent in the $LEET$.
The amplitude for the particle to remain into the classical
trajectory decays like $1/t$, so the probability decays like
$1/t^2$.

The lagrangian of the $\overline{LEET}$, omitting the spin
dependence, is
\beq\label{quarantuno}
{\cal{L}}(x)~=~Q^{\dagger}(x)\Bigg[~in\cdot D+\frac{{D_T}^2}{2E}~\Bigg]Q(x).
\eeq

We believe that the $\overline{LEET}$ is the correct effective
theory for massless particles as long as exclusive processes
are concerned.

$~~$

\centerline{\bf Acknowledgements}

$~~$

We wish to thank G. Martinelli and C. Sachrajda for discussions.

$~~$

\vfill

\end{document}